\documentclass{iopjournal}

\begin{document}

\articletype{Article type} 

\title{Erratic Liouvillian Skin Localization and Subdiffusive Transport}

\author{Stefano Longhi}

\affil{$^1$Dipartimento di Fisica, Politecnico di Milano, Piazza L. da Vinci 32, I-20133 Milano, Italy}

\affil{$^2$IFISC (UIB-CSIC), Instituto de F\'isica Interdisciplinar y Sistemas Complejos, Campus Universitat de les Illes Balears, E-07122 Palma de Mallorca, Spain}

\email{stefano.longhi@polimi.it}


\begin{abstract}
Non-Hermitian systems with globally reciprocal couplings -- such as the Hatano-Nelson model with stochastic imaginary gauge fields -- avoid the conventional non-Hermitian skin effect, displaying erratic bulk localization while retaining ballistic transport. An open question is whether similar behavior arises when non-reciprocity originates at the Liouvillian level rather than from an effective non-Hermitian Hamiltonian obtained via post-selection. Here, a lattice model with globally reciprocal Liouvillian dynamics and locally asymmetric incoherent hopping is investigated, a disordered setting in which Liouvillian-specific effects have remained largely unexplored. While the steady state again shows disorder-dependent, erratic localization without boundary accumulation, {\color{black}excitations in the incoherent-hopping regime spread via {\em Sinai-type subdiffusion}, dramatically slower than ordinary diffusion in symmetric stochastic lattices.} {\color{black}This highlights that the genuinely distinct Liouvillian signature is the coexistence of global reciprocity with ultra-slow, disorder-induced subdiffusive transport, rather than the erratic localization itself.} {\color{black}These results reveal a fundamental distinction between globally reciprocal Hamiltonian and Liouvillian systems: in both cases the skin effect is suppressed, but only in Liouvillian dynamics erratic skin localization can coexist with subdiffusive transport.}
\end{abstract}

\section{Introduction}

Non-Hermitian (NH) physics \cite{NH1} has emerged as a vibrant research frontier across diverse areas of physics, spanning condensed matter and open quantum systems, as well as classical platforms including photonic, acoustic, mechanical, and electrical lattices \cite{NH2,NH3,NH4,NH5,NH6,NH7,NH8,NH9,NH10,NH11,NH12,NH13,NH14}.
 Non-Hermiticity can fundamentally alter the spectral and dynamical properties of lattice models, as well as their topological phases \cite{NH14,NH15}. One of the most striking manifestations of these effects is the non-Hermitian skin effect (NHSE) \cite{SK1,SK2,SK3}, in which an extensive number of eigenmodes under open boundary conditions (OBC) become localized at the edges of a lattice, typically due to asymmetric (non-reciprocal) hopping. This phenomenon, together with its diverse manifestations and broad range of applications, has been extensively investigated (see, e.g., \cite{NH6,NH7,NH10,NH11,NH13,SK1,SK2,SK3,SK4,SK5,SK6,SK7,SK8,SK9,SK10,SK11,SK12,SK13,SK14,SK15,SK16,SK17,SK18,SK18b,SK19,SK20,SK21,SK22,SK23,Brunelli,SK24,SK25,SK26,SK26b,Brunelli2,SK27,SK28,SK29,SK30,SK30b,SK31,SK32,SK32b,SK33,SK34} and references therein).

In open quantum systems, non-reciprocal behavior and the associated skin effect can emerge from two distinct mechanisms. In the first scenario, the effective non-Hermitian (NH) Hamiltonian describing the conditional dynamics under post-selection captures the asymmetry in coherent hopping \cite{NH9,SK8,uff1,uff2,uff3,uff4,uff5,LS2,uff6}. A prototypical example is the Hatano-Nelson model \cite{NH14,Hatano}, in which unequal left- and right-hopping rates lead to the accumulation of eigenmodes at one lattice edge and to biased ballistic transport in the bulk, accompanied by transient self-acceleration rooted in the spectral topology of the NHSE \cite{SK28,self}. In the second scenario, non-reciprocity arises from the Liouvillian superoperator governing the full density matrix dynamics, giving rise to the so-called Liouvillian skin effect (LSE) \cite{SK8,LS2,LS1}. A possible route to realize the LSE is via asymmetric incoherent hopping terms in the dissipative dynamics \cite{LS11,LS0}, which induce the localization of right and left eigenmodes of the Liouvillian toward lattice edges and bias transport in the bulk, even when the corresponding effective NH Hamiltonian does not exhibit the NHSE \cite{LS1}. The LSE and related non-normal spectral features have been investigated in in several recent studies \cite{SK8,uff5,LS2,LS1,LS3,LS4,LS5,LS6,LS7,LS8,LS9,LS10} and linked to long transients \cite{LS1}, sensitivity to boundary conditions \cite{LS4}, chiral damping \cite{SK8}, slow or enhanced relaxation \cite{LS1,LS3}, anomalous transport and the quantum Mpemba effect \cite{LS9,LS10}.

Most previous studies have focused on ordered systems, where the lattice parameters are uniform and disorder is absent, or on disordered systems displaying an overall global non-reciprocity and the usual NHSE with a point-gap topology within an effective NH Hamiltonian description \cite{SK18b,Brunelli,SK26b,SK30b,SK32b,SK34}. Recently, however, a variant of the Hatano-Nelson model with stochastic imaginary gauge fields has been introduced, which is globally reciprocal despite local non-reciprocity \cite{SK32,T1,T2,T3}. In this globally symmetric configuration, the NHSE is suppressed: eigenmodes no longer accumulate at lattice edges under OBC \cite{SK32}, and transport remains ballistic, essentially reproducing the behavior of a Hermitian lattice with symmetric couplings \cite{T1}. This raises the natural question of how global reciprocity manifests in Liouvillian dynamics, where non-Hermitian behavior originates from dissipative processes rather than coherent hopping.
While this question provides one point of departure, the broader motivation of the present work is to uncover phenomena that are genuinely intrinsic to Liouvillian dynamics and have no analogue in effective non-Hermitian Hamiltonians obtained via post-selection. In disordered systems, this distinction becomes particularly pronounced. Yet, despite their fundamental importance, Liouvillian-specific effects in disordered, globally reciprocal settings remain scarcely explored. Characterizing the resulting spectral structure and dynamical behavior is therefore essential for understanding transport and relaxation in driven-dissipative lattices beyond the paradigms established by non-Hermitian Hamiltonian models.

In this work, we therefore analyze a prototypical bosonic lattice model of incoherent transport \cite{LS1,LS3,LS7} under globally reciprocal stochastic hopping. We show that, even though the LSE is destroyed by global symmetry, the steady state exhibits an {\em erratic, sample-specific form of localization} that is qualitatively distinct from conventional skin accumulation. More strikingly, we demonstrate that global reciprocity at the Liouvillian level does {\em not} guarantee conventional transport behavior. Instead, in the incoherent-hopping-dominated regime, excitations propagate via {\em Sinai-type subdiffusion}, spreading logarithmically slowly in time -- a dramatic departure from both the ballistic dynamics of the globally reciprocal Hatano-Nelson model and the diffusive transport expected in symmetric stochastic lattices.
These findings reveal that Liouvillian dynamics can host fundamentally different disorder-induced and reciprocity-protected phenomena compared to their non-Hermitian Hamiltonian counterparts. They establish erratic localization and Sinai-type subdiffusion as intrinsic dynamical signatures of globally reciprocal Liouvillians with incoherent hopping, thereby uncovering a new regime of transport and spectral behavior in open quantum systems.

\section{Non-Hermitian and Liouvillian skin effects in a tight-binging lattice: Model}
We consider a one-dimensional tight-binding lattice described by the Hamiltonian
\begin{equation}
    H = J \sum_n \left( a_n^\dagger a_{n+1} + a_{n+1}^\dagger a_n \right),
\end{equation}
where $a_n$ ($a_n^\dagger$) denotes particle annihilation (creation) operator at site $n$, and $J$ is the nearest-neighbor hopping amplitude. Throughout this work we assume bosonic particles. Nevertheless, because our analysis concerns primarily single-particle excitation dynamics, the main results remain valid for fermionic systems as well. The system is \emph{open} in the sense that it interacts with external baths, and its dynamics is therefore governed by a Lindblad master equation for the density operator $\rho(t)$ (see e.g. \cite{NH9,SK8,LS1})
\begin{equation}
    \frac{d \rho}{dt} = -i [H, \rho] + \sum_l \left( L_l \rho L_l^\dagger - \frac{1}{2} \{ L_l^\dagger L_l, \rho \} \right) \equiv \mathcal{L} \rho,
\end{equation}
where $L_l$ are the jump operators describing the dissipative processes induced by the environment.
For certain choices of jump operators, it is convenient to introduce an effective non-Hermitian Hamiltonian
\begin{equation}
    H_{\rm eff} = H - \frac{i}{2} \sum_l L_l^\dagger L_l,
\end{equation}
which governs the dynamics under conditional, no-jump evolution. 

\begin{figure}[t]
\centering
\includegraphics[width=0.4\textwidth]{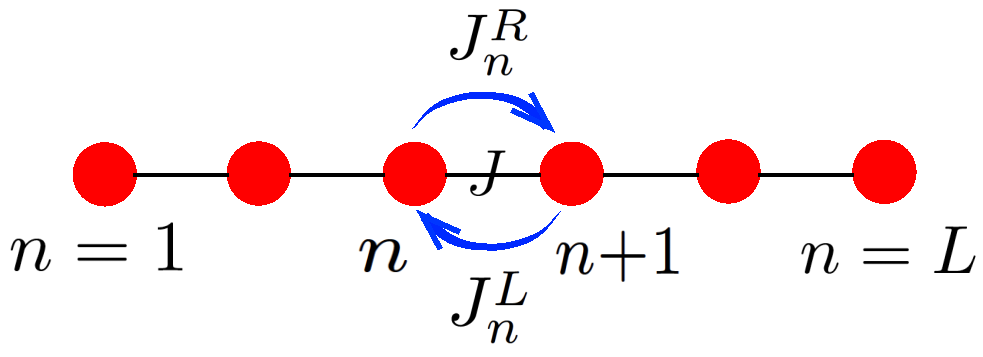}
\caption{Schematic a tight-binding lattice made of $L$ sites with coherent ($J$) and site-dependent asymmetric incoherent($J^R_n, J^L_n$)  hopping between adjacent sites. Open boundary conditions are assumed. The incoherent hopping rates are given by $J^L_n=Q \exp(h_n)$ and $J_n^R=Q \exp(-h_n)$, where $h_n$ is the asymmetry parameter. For $h_n=h \neq 0$, the model displays the LSE. When $h_n$ are independent stochastic variables that can assume only the two values $+h$ or $-h$ with the same probability, the system is globally reciprocal and the LSE disappears. Instead, erratic Liouvillian skin localization is observed, associated to sub-diffusive spreading of an initially-localized boson in the lattice bulk.}
\label{fig1}
\end{figure}

\vspace{0.5cm}
\noindent Non-reciprocity in the system, which leads to biased transport and is associated with either the NHSE or the LSE, can arise through two fundamentally distinct mechanisms.

\vspace{0.5cm}

\noindent { \em 1.  Non-Hermitian Skin Effect.}\\
\noindent The first route introduces non-reciprocity through the jump operators \cite{NH9,SK8,Brunelli,Brunelli2,uff1,uff2,uff3,uff4,uff5,uff6}
\begin{equation}
    L_n = \sqrt{\kappa} \left(a_n - i e^{i \theta_n} a_{n+1}\right),
\end{equation}
which are non-local but linear in the bosonic destruction operators. Here, $\kappa$ is the dissipation rate and $\theta_n$ a phase that controls the local asymmetry. Under conditional, no-jump evolution, the system dynamics is governed by the effective NH Hamiltonian $H_{\rm eff}$ which describes a stochastic version of the Hatano-Nelson model. This case has been studied in several previous works (see e.g. \cite{NH9,uff3,uff4,uff5,LS2,uff6}), and it briefly reviewed in the Appendix A.  In the absence of disorder, setting $\theta_n = 0$ with $ \kappa< 2J$ reduces $H_{\rm eff}$ to the clean Hatano-Nelson model,
\begin{equation}
    H_{\rm HN} = \sum_n \left\{ (J - \kappa/2) a_n^\dagger a_{n+1} + (J + \kappa/2) a_{n+1}^\dagger a_n \right\} -\frac{i}{2} \kappa \sum_n \left( a_n^\dagger a_{n} +  a_{n+1}^\dagger a_{n+1} \right)
\end{equation}
which exhibits the NHSE, i.e. a macroscopic accumulation of the eigenstates of the NH Hamiltonian at the lattice edges under open boundary conditions, due to effective asymmetric (non-reciprocal) left/right hopping amplitudes $J^{R}=J+ \kappa/2$, $J^{L}=J-\kappa/2$.  
The stochastic and globally-reciprocal Hatano-Nelson model \cite{SK30b,SK32,T1} is obtained by assuming that $\theta_n$ are independent random variables with a Bernoulli probability distribution that can assume with equal probability either one of the two values $\theta_n = 0$ or $\pi$. The localization and transport properties of this model have been deeply investigated in recent works \cite{SK30b,SK32,T1} and here we just briefly summarize the main results, which are relevant for the current work. The globally-reciprocal random Hatano-Nelson model does not display the ordinary NHSE, rather it shows disorder-dependent local stochastic NH skin interfaces \cite{SK32} where excitation locally accumulates, a phenomenon dubbed erratic NH skin localization. The stochastic interfaces at  which excitation accumulates are governed by the universal order statistics of random walks \cite{SK32,T1}. Remarkably, transport in this disordered globally-reciprocal lattice remains ballistic on average, consistent with recent observations in fluctuating globally-reciprocal NH lattices \cite{T1}. Specifically, since on average the left and right hopping amplitudes are symmetric, there is not biased transport and, on average, one observes the same ballistic spreading of excitation as in a clean Hermitian lattice with effective hopping rate ${J}_e=\sqrt{J^2- \kappa^2/4}$ \cite{T1}. In other words, restoration of global reciprocity restores the ballistic transport nature of a symmetric-coupling (Hermitian) lattice.\\ 
\vspace{0.5cm}
\noindent { \em 2. Liouvillian Skin Effect.}\\ 

\noindent A second way to introduce non-reciprocity and biased bosonic transport in the lattice is via asymmetric incoherent hopping, which has been introduced and investigated in several recent works \cite{LS11,LS0,LS1,LS3,LS4,LS5,LS6,LS7,LS8,LS9,LS10} and related to the celebrated asymmetric simple
exclusion process \cite{LS0,LS7}. The corresponding jump operators read
\begin{equation}
    L_n^R = \sqrt{J_n^R} a_{n+1}^\dagger a_n, \quad L_n^L = \sqrt{J_n^L} a_n^\dagger a_{n+1},
\end{equation}
which are quadratic in the bosonic operators. Here, $J_n^{R,L}$ are site-dependent rates for right- and left-biased incoherent hopping. Contrary to the previous case of linear jump operators, 
in this case the evolution generated by the Liouvillian $\mathcal{L}$ preserves the
total number of bosons $N=\sum_n a^{\dag}_n a_n$, which constitutes a strong symmetry of the system, {\color{black}{i.e. $N$ commutes both with the Hamiltonian and with all Lindblad jump operators} }. In the absence of disorder, homogeneous rates $J_n^R = J^R$ and $J_n^L = J^L$ produce a LSE whenever $J^R \neq J^L$, manifested as edge localization of Liouvillian eigenmodes -- rather than eigenmodes of the effective NH Hamiltonian --  and biased boson transport in the bulk \cite{LS1,LS7}. Unlike the case with jump operators linear in the bosonic fields -- where the dynamics remain Gaussian and the model is exactly solvable via the so-called "third quantization" approach (see e.g. \cite{LS2,uff6}) -- the present setup with quadratic jump operators is no longer exactly solvable rather generally \cite{LS11,LS0,LS1,LS7}. Nevertheless, valuable physical insights into the relaxation dynamics and transport can still be gained by analyzing the single-particle sector ($N=1$) \cite{LS1}. In this regime, particle statistics are irrelevant, so the model can be applied to fermions as well.\vspace{0.5cm}

\begin{figure}[t]
\centering
\includegraphics[width=0.95\textwidth]{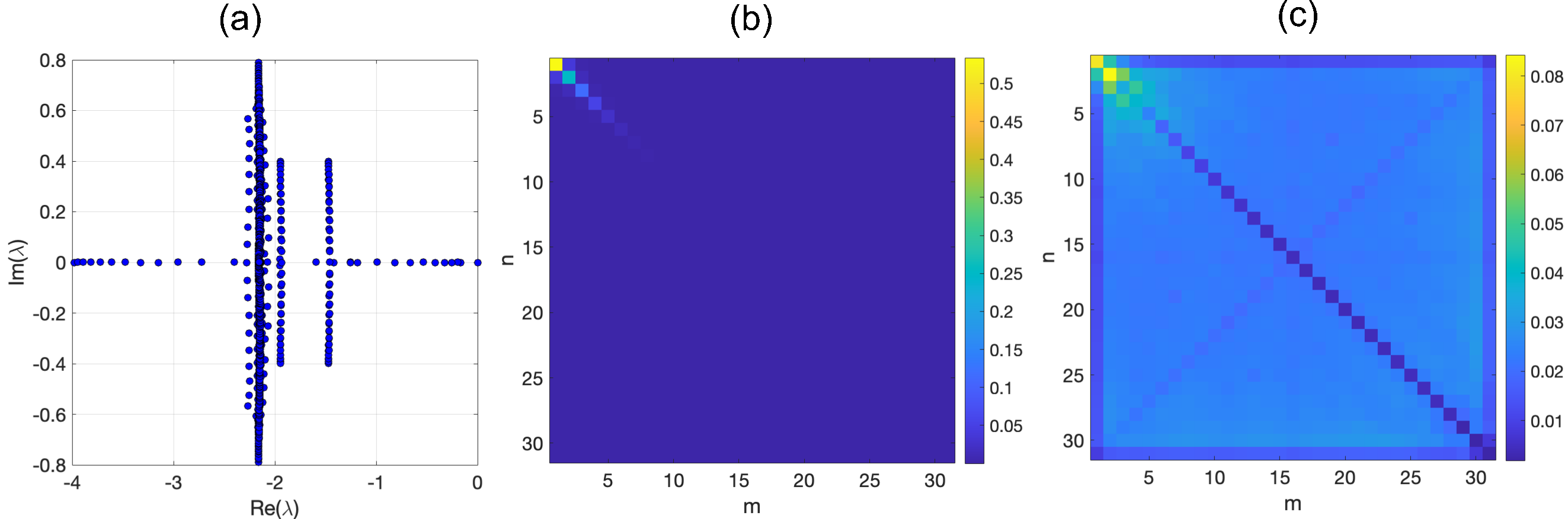}
\caption{Liouvllian skin effect in a tight-binding lattice with uniform asymmetric parameter $h_n=h$. (a) Spectrum (eigenvalues $\lambda_{\alpha}$) of the Liouvillian $\mathcal{L}$ in the single-particle sector under OBC for parameter values $J=0.2$, $Q=1$, $h=0.4$ and lattice size $L=31$. (b) Equilibrium density matrix (plot of $|\rho^e_{n,m}|$ on a pseudo-color map). (c) Distribution $I_{n,m}$ of averaged right eigenvectors of the Liouvillian $\mathcal{L}$. Note the localization of $\rho^e$ and $I_{n,m}$ on the upper left corner, a characteristic signature of the LSE.}
\label{fig2}
\end{figure}
\begin{figure}[t]
\centering
\includegraphics[width=0.95\textwidth]{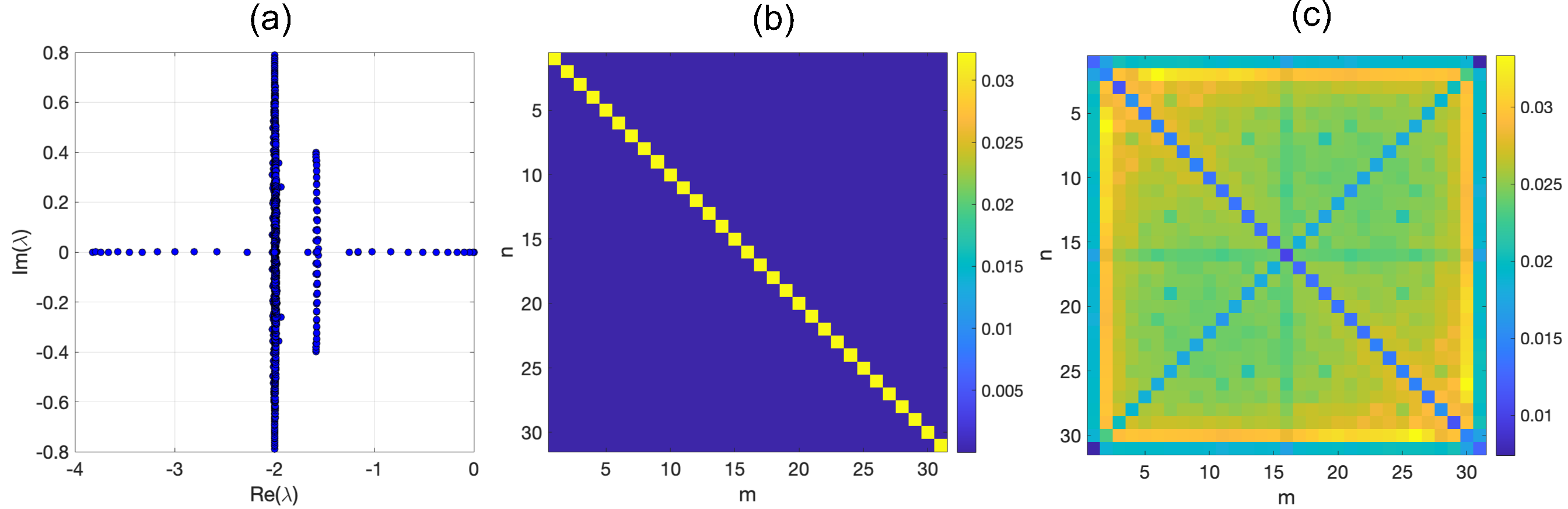}
\caption{Same as Fig.2, but for $h=0$ (reciprocal and disorder-free model). Note the disappearance of the LSE.}
\label{fig3}
\end{figure}
\begin{figure}[t]
\centering
\includegraphics[width=0.95\textwidth]{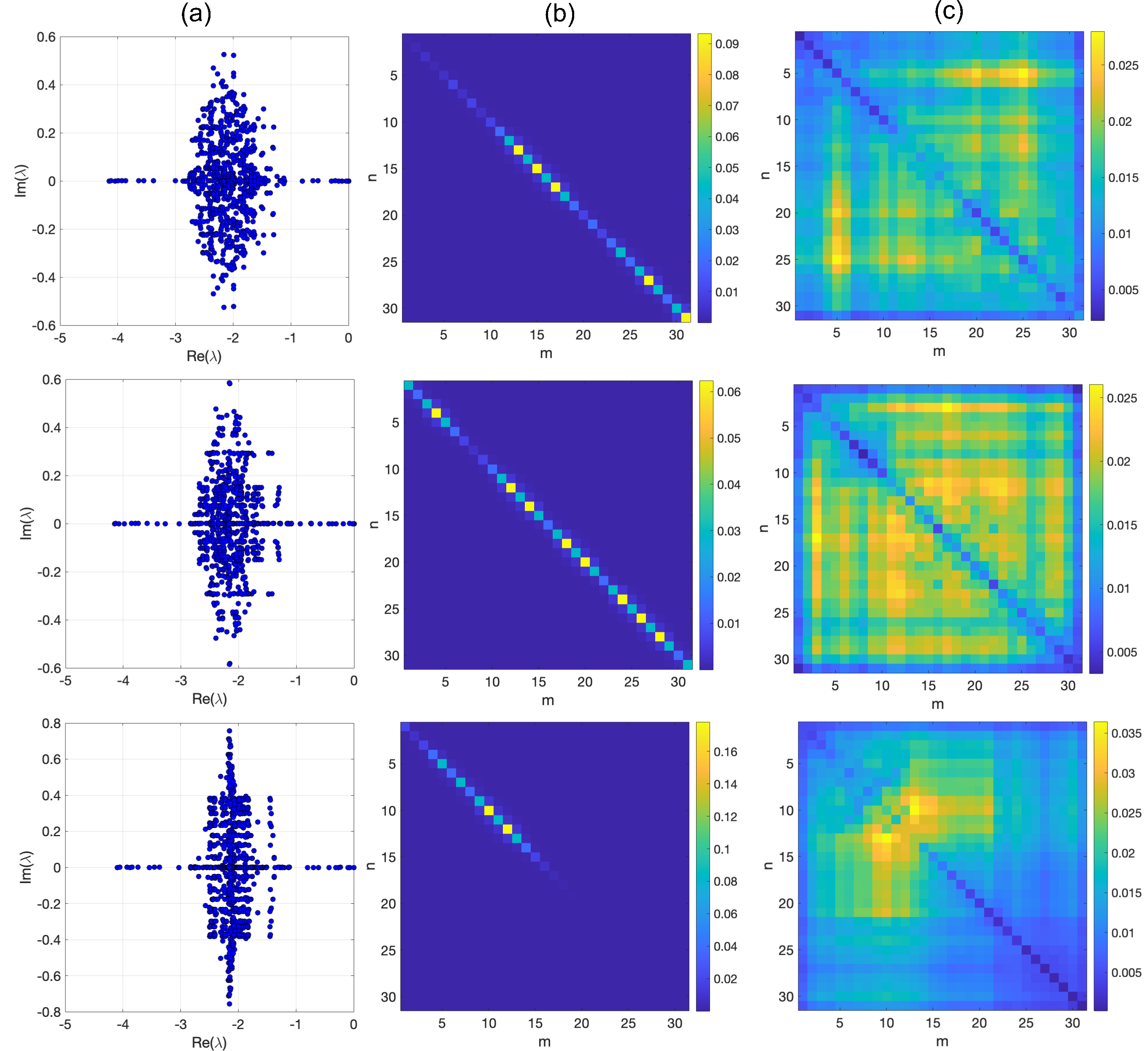}
\caption{Same as Fig.2, but in the stochastic lattice where $h_n$ can take only the two values $h_n= \pm h$ with the same probability. Parameter values are as in  Fig.2 ($J=0.2$, $Q=1$, $h=0.4$ and $L=31$). The three row in the figures refer to three realizations of the stochastic sequence $ \{h_n \}$.}
\label{fig4}
\end{figure}

In the following analysis, we focus our attention to this second mechanics of non-reciprocity, leading to the LSE for homogeneous but asymmetric hopping rates $J^{L} \neq J^R$ \cite{LS1}. Specifically, we consider  a disordered system with {\em global reciprocity}. This is obtained assuming 
\begin{equation}
J_n^{L}=Q \exp(h_n) \;, \;\; J_n^{R}=Q \exp(-h_n),  
\end{equation}
where $h_n$ are independent Bernoulli random variables taking the values $\pm h$ with equal probability.

\section{Erratic Liouvillian skin localization}
We consider the Liouvillian dynamics in a tight-binding chain consisting of $L$ sites with OBC and with dissipators defined by Eqs.(6) and (7); see Fig.1 for a schematic. To unveil the role of stochastic unbalance terms $h_n$ on the LSE and transport features in the lattice, following previous works \cite{LS1,LS9} we focus our analysis to the single-particle sector $N=1$, which captures the essential features of single-particle localization and relaxation dynamics \cite{LS1}.  Restricting to the single-excitation manifold spanned by the basis $\{|n\rangle=a_n^\dagger|0\rangle\}$ with $n=1,2,...,L$, the density matrix elements are $\rho_{mn}=\langle m|\rho|n\rangle$, and the statistical nature of the particle (i.e. either bosons or fermions) becomes irrelevant in this regime. Using the Hamiltonian~(1) and the jump operators~(6), from the Lindblad master equation $(d \rho /dt)= \mathcal{L} \rho$ and under OBC one readily obtains the following evolution equations for the density matrix elements

\begin{eqnarray}
\dot{\rho}_{n,m} & = &  i \sum_{k=1}^L \left( \rho_{n,k}H_{k,m}-H_{n,k} \rho_{k,m} \right) \nonumber \\
& + & J^R_{n-1} \delta_{n,m} (1- \delta_{n,1}) \rho_{n-1,n-1}+ J^L_n \delta_{n,m} (1-\delta_{n,L}) \rho_{n+1,n+1} \\
& - &  \frac{1}{2} \rho_{n,m} \left\{ J_n^R(1-\delta_{n,L})+J_m^R (1-\delta_{m,L})+J_{n-1}^L(1-\delta_{n,1})+J_{m-1}^L (1-\delta_{m,1})  \right\}  \nonumber
\end{eqnarray}
where 
\begin{equation}
H_{n,m}= \langle n|  H | m \rangle = \left(
\begin{array}{cccccccc}
0 & J & 0 & 0 & ... & 0 & 0 & 0 \\
J & 0 & J & 0 & ...& 0 & 0 & 0 \\
... & ... & ... & ... & ... & ... & ... & ... \\
0 & 0 & 0 & 0 & ... & J & 0 & J \\
0 & 0 & 0 & 0 & ... & 0 & J & 0
\end{array}
\right).
\end{equation}
{\color{black} are the matrix elements} of the coherent (Hermitian) Hamiltonian {\color{black} on the Hilbert space basis $|n \rangle$}. We indicate by $r^{(\alpha)}$ and $\lambda_{\alpha}$ the right eigenvectors and corresponding eigenvalues of the Liouvillian $\mathcal{L}$, i.e. $ \mathcal{L}  r^{(\alpha)} = \lambda_{\alpha} r^{(\alpha)}$ ($ \alpha=1,2,...,L^2$), ordered such that $\lambda_1=0> {\rm Re}(\lambda_2) > {\rm Re}(\lambda_3)>...$. The stationary (equilibrium) state $\rho^{e}=r^{(1)}$ corresponding to the eigenvalue $\lambda_{1}=0$ is unique \cite{LS4}. In the strongly dissipative regime $J_{n}^{R,L} \gg J$, coherences $\rho_{n,m}(t)$ decay rapidly, reducing the dynamics to a classical master equation \cite{LS1,LS7,LS9}, which allows one to obtain $\rho^{e}$ analytically in closed form. See Appendix~B for details. 
All other Liouvillian eigenvectors correspond to decaying modes with rates $|\mathrm{Re}(\lambda_{\alpha})|$. For the non-stochastic case $h_n = h$, $\mathcal{L}$ exhibits the skin effect (LSE), with eigenvectors -- including the equilibrium state -- localized toward one edge of the lattice \cite{LS1}. Fig.~2(a) illustrates this behavior, showing a representative eigenvalue spectrum $\lambda_{\alpha}$, the equilibrium distribution $\rho^e$, and the average spatial profile of the right eigenvectors,
\begin{equation}
I_{n,m} = \frac{1}{L^2} \sum_{\alpha=1}^{L^2} \bigl| r^{(\alpha)}_{n,m} \bigr|.
\end{equation}
{\color{black} Physically, the quantity $I_{n,m} $ provides a global measure of the spatial localization of all Liouvillian eigenmodes. 
It captures not only the stationary state $\rho^e = r^{(1)}$ but also all decaying modes, weighted equally, thus highlighting lattice sites where the eigenmodes tend to accumulate. 
In this way, $I_{n,m}$ gives a comprehensive overview of the Liouvillian skin effect, showing how both equilibrium and transient dynamics are spatially biased due to the non-reciprocal or disordered hopping. 
This measure is commonly used in the literature to visualize eigenmode boundary localization in non-Hermitian and Liouvillian systems \cite{SK24}.
As shown in Fig.3, } the localization edge changes from left to right as the sign of $h$ is flipped, while the LSE disappears in the reciprocal limit $h_n=0$. \\
\vspace{0.2cm}
\noindent In the disordered and globally-reciprocal system, where each $h_n$ independently takes one of the two values $+h$ or $-h$ with equal probability, the LSE--that is, the boundary accumulation of the equilibrium state and of other Liouvillian eigenvectors -- no longer occurs. Instead, one typically observes a disorder-dependent bulk localization, often exhibiting one or several irregular peaks, closely resembling the \emph{erratic skin localization} recently introduced in Ref.~\cite{SK32}. This behavior is illustrated in Fig.~4. We refer to this disorder-dependent, irregular localization of the stationary state $\rho^e$ as \emph{erratic Liouvillian skin localization}.  
The disappearance of the LSE and the onset of erratic Liouvillian skin localization for the equilibrium state, observed in numerical simulations, can be rigorously understood in the fully dissipative limit $J=0$, where the dynamics reduces to a classical master equation for the populations $P_n(t) = \rho_{n,n}(t)$ (see Appendix B). Specifically, the equilibrium populations $P^e_n$ are maximally localized at lattice sites corresponding to the extreme values of the cumulative random variable
\begin{equation}
X_n = \sum_{l=1}^{n-1} h_l, \quad X_1 = 0.
\end{equation}
Since each $h_n$ takes $+h$ or $-h$ with equal probability, $X_n$ represents a symmetric random walk on the line. {\color{black} Consequently, the equilibrium-state localization is strongly disorder-dependent and closely parallels the erratic skin localization previously predicted in Ref.~\cite{SK32} for the globally reciprocal Hatano-Nelson model.}

\section{Sub-ballistic Sinai transport}

In this section, we investigate the nature of Liouvillian transport in the lattice, focusing on the regime where dynamics is dominated by the dissipative (incoherent) hopping terms ($J \ll Q$). We consider an initial condition where a single particle is localized at the middle site of the lattice, $n=n_0=(L+1)/2$ (assuming $L$ odd). The relaxation dynamics drives the system toward the unique stationary state $\rho^e$. Our goal is to explore the {\it bulk dynamics}, i.e., the spreading of excitation in the thermodynamic limit $L \rightarrow \infty$ where edge effects are negligible.
{\color{black}  Owing to the stochastic nature of ${h_n}$, the identification of localization features and transport regimes requires ensemble averaging, since a single disorder realization, while capturing qualitative mechanisms, does not provide reliable access to universal spreading laws. Controlled disorder and ensemble averaging are standard in reconfigurable experimental platforms, such as photonic, atom-optics, or acoustic systems, where local parameters can be programmatically varied to sample multiple disorder configurations (see e.g. Ref.\cite{T1}).}

 Excitation spreading in the lattice is characterized by the mean particle position $n_{CM}(t)$ and the second moment $d^2(t)$, defined as
\begin{equation}
n_{CM}(t)=\sum_n (n-n_0) \rho_{n,n}(t) \; , \;\;\; d^2(t)=\sum_n (n-n_{CM})^2 \rho_{n,n}(t).
\end{equation}
When $h_n=h$ is uniform and nonzero, the system exhibits the LSE. As expected, the excitation drifts along the lattice with a constant speed and diffusive spreading around the center of mass, namely
\begin{equation}
n_{CM}(t) \propto t \; , \;\;\; d^2(t) \propto t,
\end{equation}
see Fig.~5 for an illustrative example.

\begin{figure}[t]
\centering
\includegraphics[width=0.95\textwidth]{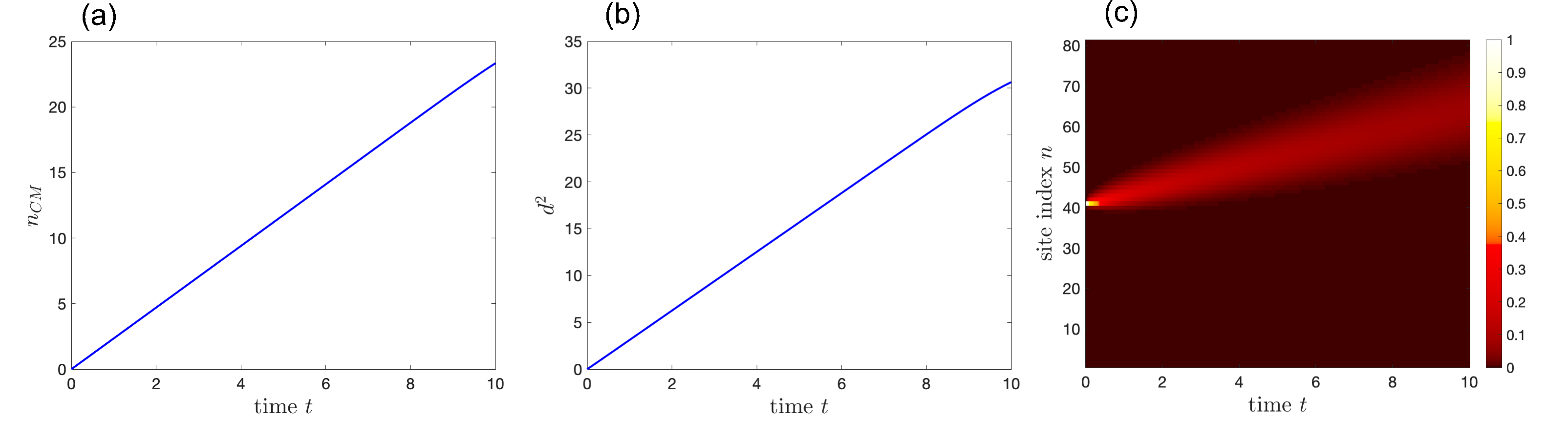}
\caption{Excitation spreading in a lattice with uniform asymmetry parameter $h_n=h$. (a,b) Numerically-computed temporal behavior of the excitation center of mass $n_{CM}(t)$ [panel (a)] and second-moment $d^2(t)$ [panel (b)]. (c) Excitation spreading dynamics (plot of $\rho_{n,n}(t)$ on a pseudo-color map).  Parameter values are $J=0.2$, $Q=1$ and $h=1$. Lattice size $L=81$.}
\label{fig5}
\end{figure}

\begin{figure}[t]
\centering
\includegraphics[width=0.95\textwidth]{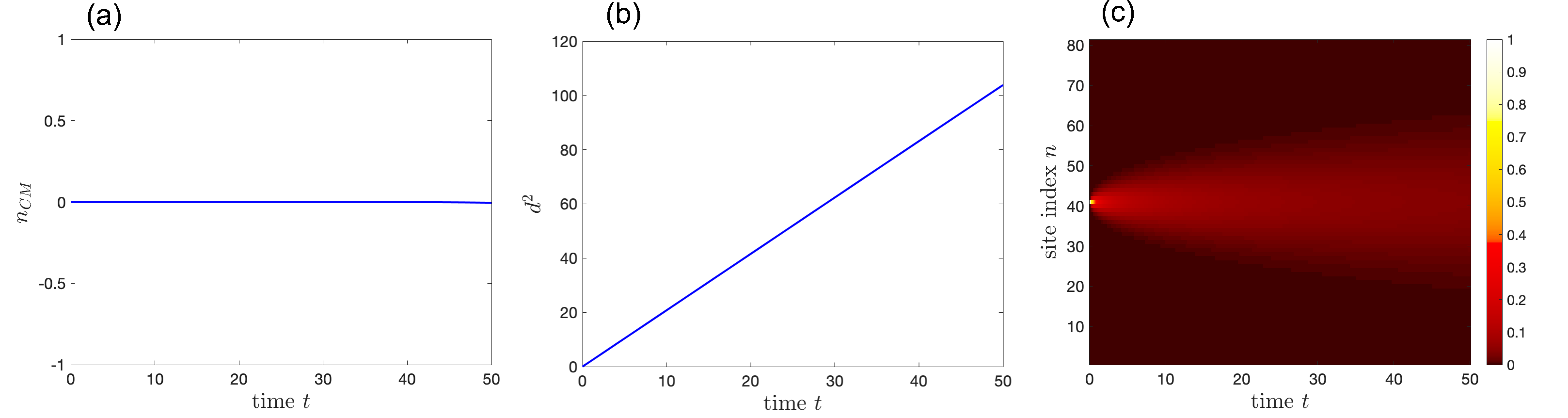}
\caption{Same as Fig.~5, but in the absence of the LSE ($h=0$).}
\label{fig6}
\end{figure}

\noindent When $h_n=0$, i.e., incoherent left/right hopping is symmetric, the mean particle position remains locked at its initial site, $n_{CM}(t)=0$, while the second moment grows linearly with time,
\begin{equation}
d^2(t) \propto t,
\end{equation}
indicating standard diffusive spreading of the excitation in the dissipative regime; see Fig.~6. A similar transport regime also appears in a disordered configuration that maintains a global bias, achieved when $h_n$ randomly takes the values $\pm h$ with unequal probabilities (see Fig.~7).

The most intriguing behavior emerges in a {\em disordered but globally reciprocal lattice}, e.g., when $h_n$ takes the values $+h$ or $-h$ with equal probability. In this case, the mean particle position remains on average near the initial site $n_0$, while the second moment $\langle d^2(t) \rangle$ grows slower than linearly, signaling {\em sub-ballistic transport}; see Fig.~8 for an illustrative example.

To gain physical insight into the observed behavior, it is useful to examine the {\em classical limit} $J=0$ of purely dissipative dynamics, where the evolution is fully captured by the diagonal elements of the density matrix. In this limit, the populations $P_n(t)=\rho_{n,n}(t)$ obey the classical master equation (Appendix~A):

\begin{equation}
\frac{d P_n}{dt} = J_{n-1}^R P_{n-1} + J_n^L P_{n+1} - (J_n^R + J_{n-1}^L) P_n,
\end{equation}
with $J_n^R = Q e^{-h_n}$ and $J_n^L = Q e^{h_n}$, where $h_n$ are independent random variables taking values $\pm h$ with equal probability.  
Because the {\em local bias} $\ln(J_n^R/J_n^L)$ is a zero-mean random variable, the dynamics maps onto a classical Sinai random walk \cite{Sinai1,Sinai2,Sinai3,Sinai4}. In this case, the transport is subdiffusive, with the typical displacement growing extremely slowly, and the long-time asymptotic behavior is predicted to follow \cite{Sinai1,Sinai2}:
\begin{equation}
\langle d^2(t) \rangle \sim (\log t)^4.
\end{equation}
The mean displacement $n_{CM}(t)$ remains on average essentially zero, reflecting the absence of a global bias.  
Physically, the subdiffusive behavior observed in the Sinai regime arises from the interplay between disorder and local hopping asymmetry. Each site $n$ is associated with a local bias (random force)  $F_n=\ln(J_n^R/J_n^L)$, which can be interpreted as a random potential landscape with ``valleys'' and ``barriers'' \cite{Sinai1,Sinai2}. Excitations tend to become temporarily trapped in deep valleys, where the local bias opposes motion, leading to long waiting times before escape. The absence of a net global bias ensures that these traps are distributed symmetrically, so that on average the mean position $\langle n_{CM}(t) \rangle$ remains near the initial site. Transport is dominated by rare, extremely slow hops over high barriers, producing the characteristic logarithmic-in-time growth of the second moment, $\langle d^2(t) \rangle \sim (\log t)^4$. This scenario contrasts with standard diffusion, where transport is controlled by short, uncorrelated hops; here, long-lived local traps induce strong correlations in the dynamics. Intuitively, one can view the excitation as performing a random walk in a rugged landscape, where the probability of escaping a deep valley decreases exponentially with its depth, leading to dramatically slowed dynamics at long times. This picture also provides a clear explanation of why disorder with zero mean produces sub-ballistic transport: while locally the excitation moves forward or backward, globally the absence of a net slope prevents ballistic drift, and transport is controlled by the extreme-value statistics of the deepest potential valleys.
To confirm the asymptotic subdiffusive regime, we numerically integrated the classical rate equations (15) on a wide lattice for long times. The results, shown in Fig.~9, exhibit excellent agreement between the theoretical predictions and numerical curves, confirming the expected logarithmic-slow growth characteristic of Sinai transport.

\begin{figure}[t]
\centering
\includegraphics[width=0.95\textwidth]{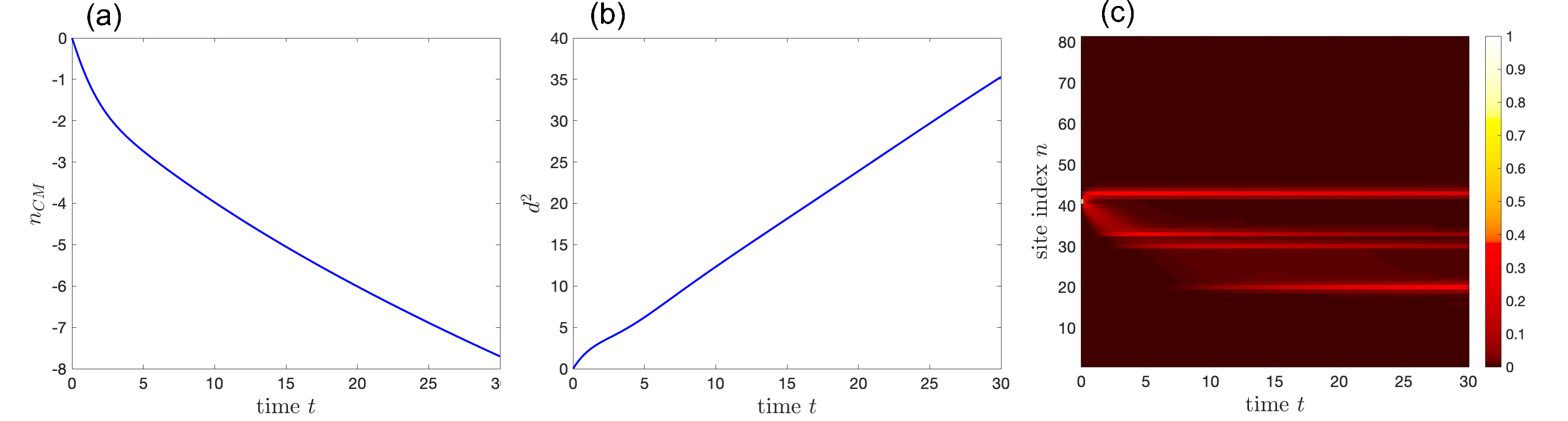}
\caption{Same as Fig.~5, but for a disordered lattice with local and global non-reciprocity. The asymmetry parameter $h_n$ can take either one of the two values $h_n=h$ or $h_n=-h$ with probabilities $p=0.3$ and $(1-p)=0.7$, respectively. The curves in (a) and (b) are obtained after averaging over 100 realizations of the random sequences $\{ h_n \}$, whereas the excitation spreading dynamics in (c) corresponds to a single disorder realization.}
\label{fig7}
\end{figure}

\begin{figure}[t]
\centering
\includegraphics[width=0.95\textwidth]{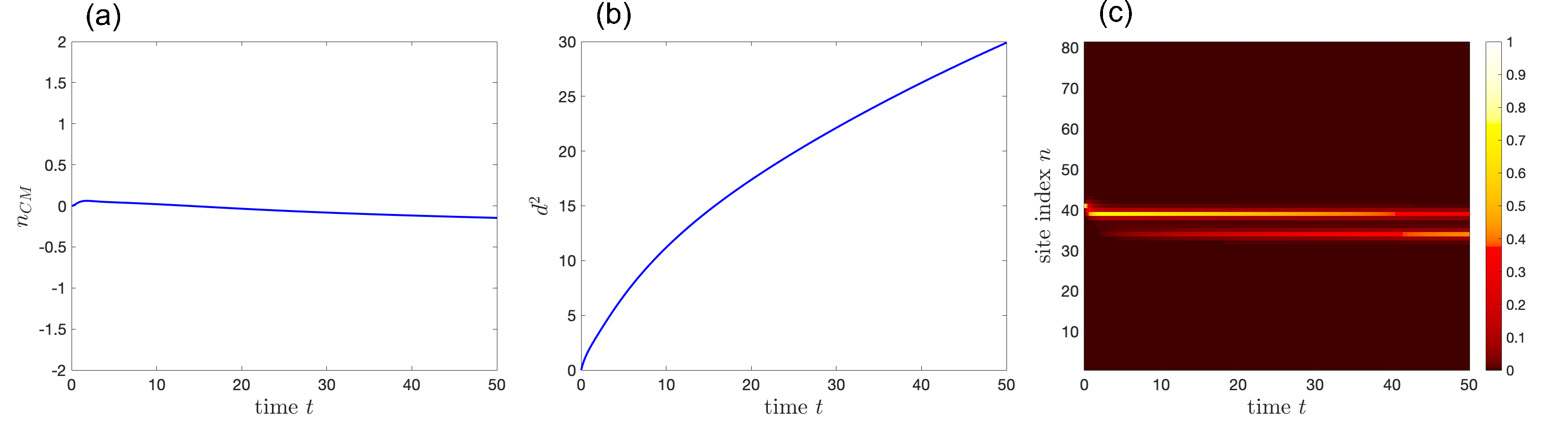}
\caption{Same as Fig.~7, but for a disordered and globally-reciprocal lattice ($h_n=h$ or $h_n=-h$ with the same probability). The curves in (a) and (b) are obtained after averaging over 100 realizations of the random sequences $\{ h_n \}$, whereas the excitation spreading dynamics in (c) corresponds to a single disorder realization, clearly showing trapping effects.}
\label{fig8}
\end{figure}

\begin{figure}
\centering
\includegraphics[width=0.95\textwidth]{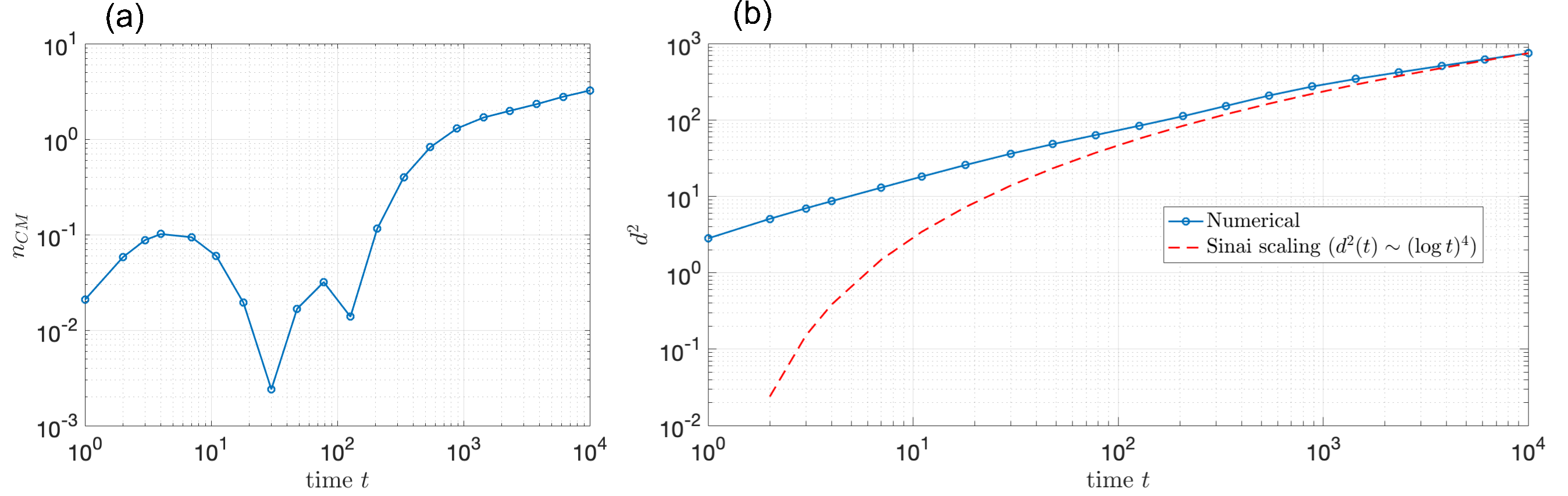}
\caption{Temporal behavior of (a) mean position $n_{CM}(t)$ and (b) the second moment $d^2(t)$ (blue curves) on a log-log scale, obtained by numerical solution of the classical master equation (15) for parameter values $Q=1$ and $h=1$. Lattice size $L=401$. The curves are obtained after averaging over 1000 realizations of the stochastic sequence $\{ h_n \}$. In (b) the red curve shows, for comparison, the Sinai scaling $d^2(t) \sim (\log t)^4$ of subdiffusion.}
\label{fig9}
\end{figure}

{\color{black} The effects of a small coherent hopping $J$ can be included in the analysis by asymptotic methods assuming $\epsilon= J/Q$ as a small parameter, as shown in the Appendix B.  In this limit, coherent processes generate coherence-assisted hopping terms that enter the population rate equations. More precisely, the effective left/right hopping amplitudes $\tilde{J}_n^{L,R}$ in the rate equations (15) acquire additional contributions of order $\epsilon^2$ from the unperturbed values $J_n^{L,R}=Q \exp( \pm h_n)$ at $J=0$; see Eqs.(42-44) in Appendix B. Rather generally, these perturbations may introduce an important qualitative consequence on transport:
they can break the exact zero-mean condition for the local force $F_n={\rm ln} (\tilde{J}_n^{R}/\tilde{J}_n^{L})$,  which is the condition responsible for Sinai subdiffusion in the purely classical setting \cite{Sinai2}. When the coherence-induced correction  to the hopping rates yields a non-vanishing bias, the asymptotic transport regime is qualitatively characterized by the competition between
the convective term induced by the net bias and the slowing-down effect of the very large fluctuations of
the potential. These high barriers act as trapping regions between which the motion is convective \cite{Sinai2}. More specifically,
as shown in \cite{Sinai2,Sinai4} distinct regimes arise, depending on the bias strength $\mu= \overline{F_n} \sim \epsilon^2$. For zero bias ($J=0$), the system exhibits Sinai diffusion, characterized by extremely slow, logarithmic growth of displacement [Eq.(16)]. For weak bias, the particle shows anomalous drift, with sub-linear growth of the average displacement. The crossover from Sinai-like subdiffusion to biased drift occurs at a time $t^* \sim \exp( {\rm const / |\mu|})$ \cite{Sinai4}, which becomes extremely long in the $\epsilon \rightarrow 0$ limit and thus the system appears Sinai-like over very long times. 
 For intermediate bias, anomalous dispersion is predicted, with nontrivial scaling of the variance, while at further larger bias normal diffusion is recovered \cite{Sinai2,Sinai4}. However, it should be noted that, whenever the probability density function $f(h)$ of the independent and equally-distributed random variables $\{ h_n \}$ has a zero mean and is symmetric about $h=0$, i.e. for $f(-h)= f(h)$, the coherence-induced corrections to the hopping rates do not break the zero mean condition $\overline{F_n}=0$, and thus a weak coherence does not wash out the subdiffusive regime.}
 
Finally, it is worth noting that, although the present discussion focuses on
single-particle dynamics, these insights provide a conceptual framework for
understanding sub-ballistic transport in the many-particle bosonic model in the
dilute limit where the mean occupation per site is small
($\langle a_n^\dagger a_n \rangle \ll 1$). In the many-particle case, a
mean-field analysis~\cite{LS7} suggests that subdiffusive behavior persists as
long as the lattice remains dilute and nonlinear hopping effects are negligible
(see Appendix~C for technical details).
{\color{black}Moreover, the strong trapping mechanism characteristic of
Sinai-type dynamics---involving broad fluctuations of effective barriers and
extremely slow escape processes -- has parallels with dynamical slowdowns 
in interacting disordered systems. In this sense, the single-particle picture 
developed here might offer a useful starting point for interpreting the emergence 
of glassy relaxation or maby-body-localization-like subdiffusive transport in denser interacting 
regimes, even though a full many-body analysis lies beyond the scope of the 
present work.}

\section{Discussion and Conclusion}

Our work demonstrates that globally reciprocal open quantum lattices with locally asymmetric Liouvillian hopping display a strikingly different transport behavior from their non-Hermitian Hamiltonian counterparts. While global reciprocity in non-Hermitian Hamiltonians suppresses the conventional NHSE and preserves ballistic transport \cite{SK32,T1}, we have shown that global reciprocity at the Liouvillian level does \emph{not} protect transport. Instead, locally asymmetric dissipative dynamics generates a complex interplay between disorder, incoherence, and non-reciprocal local couplings, {\color{black} leading to a pronounced slowing down of bulk transport coexisting with erratic localization of the stationary state.
While erratic skin localization of the non-equilibrium steady state is fully analogous to what observed in non-Hermitian models \cite{SK32}, in the Liouvillian dynamics transport exhibits a completely different behavior, namely {\em Sinai-type subdiffusion} is observed in the incoherent-hopping-dominated (dissipative) regime}. Here, the mean displacement effectively frozen and the second moment growing extremely slowly, consistent with the logarithmic-in-time scaling predicted for classical Sinai random walks \cite{Sinai1,Sinai2,Sinai3}. Physically, this subdiffusive behavior can be understood as the consequence of a rugged, locally biased energy landscape induced by the asymmetric Liouvillian hopping: excitations become transiently trapped in regions where local biases oppose motion, leading to rare-event-dominated transport and long relaxation times. Unlike standard diffusive spreading, transport here is controlled by exponentially large trapping times and extreme fluctuations in the local hopping rates. This insight establishes a direct link between microscopic local Liouvillian asymmetries and emergent anomalous transport in open quantum lattices. {\color{black} From an experimental perspective, the ingredients of the model
presenteed in this work are well within reach of current platforms. In ultracold atoms, asymmetric
incoherent hopping with tunable and site-dependent rates $J_n^L$ and $J_n^R$ can be engineered using laser-assisted tunnelling
combined with directional optical pumping \cite{
LS1,LS7}. Synthetic photonic lattices provide another promising route, where dissipative
coupling and controlled asymmetry can be introduced using optical loss engineering or electro-optic
modulation in synthetic dimensions \cite{SK31}. Finally, a quantum-optical platform to realize Liouvillian dynamics in bosonic systems with incoherent asymmetric hopping has been recently proposed in Ref.\cite{LS8}.}

Our findings carry several important conceptual and practical implications. First, they highlight a fundamental distinction between global reciprocity in Hamiltonian versus Liouvillian systems: whereas global symmetry in the Hamiltonian ensures robust ballistic transport even in the presence of local asymmetry, global symmetry in the Liouvillian does not generically protect transport, and local asymmetries can qualitatively reshape spreading. Second, they point to the erratic Liouvillian skin localization as a subtle mechanism for controlling not only steady-state localization but also dynamical transport in open quantum systems. By engineering local asymmetries, it is possible to access regimes of extremely slow dynamics, subdiffusion, and long-lived metastable states \cite{D1}, which may be relevant for quantum memories, slow light, or synthetic dissipative materials.

Looking ahead, our results suggest several promising directions. A natural extension is the many-particle regime, where one may expect emergent hydrodynamic transport \cite{LS7}. It is equally interesting to investigate interacting many-body systems \cite{manybody0,manybody0a,manybody0b,manybody1c,manybody1b,manybody1,manybody2} governed by globally reciprocal yet locally asymmetric Liouvillians \cite{LS6}, where subdiffusive dynamics may compete with interaction-driven relaxation and many-body localization. Non-reciprocal settings, such as the non-Hermitian Kitaev chain, offer further opportunities, with pairing capable of inducing unconventional phase transitions \cite{manybody2}. Finally, the connection to classical Sinai diffusion suggests that methods from disordered statistical mechanics may yield predictive insights into open-quantum transport, motivating analytical tools for characterizing rare-event-dominated dynamics in Liouvillian systems.

\section*{Acknowledgments}
The author acknowledges the Agencia Estatal de Investigacion (MDM-2017-0711).

\appendix

\section{The stochastic non-Hermitian Hatano-Nelson model.}
The stochastic NH Hatano-Nelson model, considered in recent works \cite{SK30b,SK32,T1}, is obtained in the open quantum system context by assuming the jump operators  \cite{NH9,SK8,Brunelli,Brunelli2,uff1,uff2,uff3,uff4,uff5,manybody1c}
\begin{equation}
    L_n = \sqrt{\kappa} \left(a_n - i e^{i \theta_n} a_{n+1}\right),
\end{equation}
which are non-local but linear in the bosonic destruction operators. Here, $\kappa$ is the dissipation rate and $\theta_n$ a phase that controls the local asymmetry. Under conditional, no-jump evolution, the system dynamics is governed by the effective NH Hamiltonian
\begin{equation}
    H_{\rm eff} = H - \frac{i}{2} \sum_n L_n^\dagger L_n \equiv \sum_{n,m} (\mathcal{H}_{\rm eff})_{n,m} a^\dagger_na_m
\end{equation}
which also determines the evolution of the single-particle correlations $C_{n,m}(t) = \mathrm{Tr}( \rho(t) a_n^\dagger a_m)$ in the full dissipative dynamics according to the equation \cite{uff5,D1}
\begin{equation}
i \frac{dC_{n,m}}{dt}= \sum_l \left\{    (\mathcal{H}_{\rm eff})_{m,l}C_{n,l} -(\mathcal{H}_{\rm eff}^\dagger)_{l,n} C_{l,m}     \right).
\end{equation}
The effective non-Hermitian Hamiltonian reads explicitly
\begin{equation}
H_{\rm eff}= \sum_n \left( J_n^{(R)}  a_{n+1}^\dagger a_n+ J_n^{(L)} a_n^\dagger a_{n+1} \right)- \frac{i}{2} \kappa \sum_n \left( a_{n}^\dagger a_n+  a_{n+1}^\dagger a_{n+1} \right),
\end{equation}
where
\begin{equation}
J_n^{(R)} = J + \frac{\kappa}{2} \exp(-i \theta_n) \; , \; \; J_n^{(L)} = J - \frac{\kappa}{2} \exp(i \theta_n) \
\end{equation}
are the effective right/left hopping rates.
 In the absence of disorder, setting $\theta_n = 0$ with $ \kappa< 2J$ reduces $H_{\rm eff}$ to the clean Hatano-Nelson model,
\begin{equation}
    H_{\rm HN} = \sum_n \left\{ (J - \kappa/2) a_n^\dagger a_{n+1} + (J + \kappa/2) a_{n+1}^\dagger a_n \right\} -\frac{i}{2} \kappa \sum_n \left( a_n^\dagger a_{n} +  a_{n+1}^\dagger a_{n+1} \right)
\end{equation}
which exhibits the NHSE, i.e. a macroscopic accumulation of the eigenstates of $(\mathcal{H}_{\rm eff})$ at the lattice edges under open boundary conditions, due to effective asymmetric (non-reciprocal) left/right hopping amplitudes $J^{R}=J+ \kappa/2$, $J^{L}=J-\kappa/2$.

 The stochastic and globally-reciprocal Hatano-Nelson model is obtained by assuming $\theta_n$ independent random variables with a Bernoulli probability distribution, namely $\theta_n$ can take only the two values $\theta_n=0$ or $ \pi$ with the same probability. In this case, the left/right hopping amplitudes are stochastically set as $J \pm \kappa/2 \equiv J_e \exp( \pm h)$ and $J \mp \kappa/2 \equiv J_e \exp( \mp h)$, leading to global reciprocity. Contrary to the clean Hatano-Nelson model, its stochastic version with global reciprocity does not display the NHSE, rather  macroscopic eigenstate localization at irregular,
disorder-dependent positions is observed at stochastic interfaces governed
by the universal order statistics of random walks \cite{SK32}. On average, transport in the stochastic Hatano-Nelson model remains ballistic as in a Hermitian lattice with effective hopping amplitude $J_e=\sqrt{J^2- \kappa^2/4}$ \cite{T1}.

\section{Liouvillian dynamics in the dissipative regime: Classical master equation}

\subsection{Purely dissipative regime}
In the purely dissipative limit \(J=0\), coherences rapidly decay to zero and the dynamics is entirely described by 
the site populations \(P_n(t)=\rho_{n,n}(t)\), which obey the classical master equation
\begin{equation}
\frac{dP_n}{dt}
=
J_{n-1}^R P_{\,n-1}
+
J_{n}^L P_{\,n+1}
-
\left( J_n^R + J_{n-1}^L \right) P_n ,
\qquad n=1,\ldots,L ,
\label{eq:MasterEq}
\end{equation}
with the boundary conventions \(J_0^{R}=J_0^{L}=0\) and \(J_L^{R}=J_L^{L}=0\).
The master equation can be written in vector form
\begin{equation}
\frac{d\mathbf{P}}{dt} = G \mathbf{P},
\end{equation}
where \(\mathbf{P} = (P_1,\ldots,P_L)^T\), and the Markov generator \(G\) is
\begin{equation}
G =
\left(
\begin{array}{cccccc}
- J_1^{R}           
    & J_1^{L} 
        & 0 
            & \cdots 
                & 0 
                    & 0 \\[6pt]
J_1^{R} 
    & -\left( J_2^{R} + J_1^{L} \right) 
        & J_2^{L} 
            & \cdots 
                & 0 
                    & 0 \\[6pt]
0 
    & J_2^{R} 
        & -\left( J_3^{R} + J_2^{L} \right) 
            & \cdots 
                & 0 
                    & 0 \\[6pt]
\vdots 
    & \vdots 
        & \vdots 
            & \ddots 
                & \vdots 
                    & \vdots \\[6pt]
0 
    & 0 
        & 0 
            & \cdots 
                & - \left( J_{L-1}^{R} + J_{L-2}^{L} \right)
                    & J_{L-1}^{L} \\[6pt]
0 
    & 0 
        & 0 
            & \cdots 
                & J_{L-1}^{R} 
                    & - J_{L-1}^{L}
\end{array}
\right).
\label{eq:Gmatrix}
\end{equation}

\noindent
{\em Equilibrium state from detailed balance.}\\
\noindent
The stationary populations $\mathbf{P}_e$ of the classical master equation satisfy detailed balance, 
\begin{equation}
J_n^R P_n^e = J_n^L P_{n+1}^e, \qquad n=1,\dots,L-1,
\end{equation}
which ensures that the net probability flux between neighboring sites vanishes. Using
\begin{equation}
J_n^R = Q \, e^{-h_n}, \qquad J_n^L = Q \, e^{h_n},
\end{equation}
the recursion relation for the stationary populations reads
\begin{equation}
P_{n+1}^e = e^{-2 h_n} \, P_n^e, \qquad n=1,\dots,L-1.
\end{equation}
Starting from $P_1^e$, we obtain
\begin{equation}
P_n^e = P_1^e \, \exp\Big[-2  \sum_{l=1}^{n-1} h_l \Big], \qquad n=2,\dots,L,
\end{equation}
with normalization condition $\sum_{n=1}^L P_n^e = 1$. 
Therefore, the stationary distribution of populations is given by
\begin{equation}
P_n^e = \frac{\exp(-2X_n)}{{\sum_{m=1}^L} \exp(-2X_m)}
\end{equation}
where we have set
\begin{equation}
X_n= \sum_{l=1}^{n-1} h_l
\end{equation}
with $X_1=0$. Note that $X_n$ describes a symmetric random walk on the line when the random variable $h_n$ has zero mean. 
The equilibrium distribution $P_n^e$ displays a characteristic disorder-dependent erratic localization in the bulk defined by the extreme values of the symmetric random walk $X_n$ \cite{SK32}, as schematically shown in Fig.4(b) of the main text. On the other hand, in a biased random walk, $P_n^e$ would display exponential localization toward one of the two lattice edges, i.e. the LSE [Fig.2(b) of the main text].

\vspace{0.5 cm}
\noindent
{\em Absence of the LSE}\\
\noindent
We define a diagonal similarity transformation
\begin{equation}
U = \mathrm{diag}(e^{-X_1}, e^{-X_2}, \dots, e^{-X_L})
\end{equation}
and the transformed matrix
\begin{equation}
W = U^{-1} G U.
\end{equation}
Then $W$ is symmetric (Hermitian) with elements
\begin{equation}
W_{n,n} = G_{n,n}, \quad
W_{n,n+1} = W_{n+1,n} = \sqrt{J_n^L J_n^R}=Q.
\end{equation}
Hence $G$ is similar to the Hermitian matrix $W$, and they share the same eigenvalues and with eigenvectors obtained one another applying the similarity transformation $U$. Since $W$ is Hermitian, its eigenvectors (including the equilibrium stationary state) do not display the skin effect, i.e. a macroscopic localization toward the lattice edges. Since the random walk $X_n$ is symmetric and the eigenvectors of $G$ and $W$ are obtained one another by the multiplication factor $\exp(-X_n)$ at each site $n$, the eigenvectors of $G$ do not display the skin effect neither. 

{\color{black}
\subsection{Dissipative-dominated regime}
Let us consider the bulk dynamics in a one-dimensional lattice in the large $L$ limit, so that we can neglect boundary effects.
When the dynamics include the coherent Hamiltonian part $H$, the evolution equation for the populations $P_n(t)= \rho_{n,n}(t)$ [Eq.(8) in the main text] reads
\begin{equation}
\frac{dP_n}{dt}=-2J {\rm Im} ( \rho_{n,n+1}) +2J {\rm Im} ( \rho_{n-1,n})+J_{n-1}^RP_{n-1}+J_n^LP_{n+1}-(J_n^R+J_{n-1}^L)P_n
\end{equation}
which deviates from the classical rate equation (23) owing to the contribution from the coherences  $\rho_{n,n+1}(t)$ and $\rho_{n-1,n}(t)$. 
Clearly, for a non-vanishing yet small coherent hopping rate $J$ ($J \ll J_n^L,J_n^R$) the leading dynamics is dominated by dissipative incoherent hopping on the lattice.
This classical population dynamics  occurs mainly on two different time scales \cite{timescale}: an initial fast dynamics followed by a slow evolution, related to the Liouvillian eigenmodes with large and small damping rates. Such a  separation of time scales  in the classical dynamics can be explained observing that, once a particle enters a valley of the random potential, the local left/right rates strongly prefer motion toward the local minimum. As a result, on a relatively short time scale (polynomial in valley size and of order $ \sim 1/Q$) the population distribution $P_n$ quickly relaxes to a quasi-stationary Gibbs-like profile around the local minimum \cite{timescale}.
In this stage, the particle essentially "thermalizes" within that valley. The population motion across major potential barriers then occurs on a much longer time scale \cite{timescale}. It is precisely on such a longer time scale that non-vanishing coherences $\rho_{n,n+1}(t)$ and $\rho_{n-1,n}(t)$ can modify transport features in the lattice.  As we will show below, the long-time asymptotic evolution of the populations can be still described by an effective rate equation model like in Eq.(23), but with effective left/right incoherent hopping rates $\tilde{J}_{n}^{L,R}$ perturbed from the values $J_n^{L,R}=Q \exp( \pm h_n)$ owing to a coherent-hopping assisted contribution. To prove this statement, let us assume $\epsilon=J/Q \ll 1$, so that the dynamical equations of the density matrix elements [Eq.(8) in the main text] can be approximately solved using asymptotic methods \cite{pert1,pert2}. The equations for the coherences $\rho_{n,m}$ with $n \neq m$ clearly indicate that, after an initial fast relaxation transient occurring on the fast time scale $ \sim 1 /Q$, they  are damped and take small values, at least of order $\epsilon$. Specifically, the leading order terms of coherences, of order $ \sim \epsilon$, are $\rho_{n,m}$ with $|n-m|=1$, while coherence terms $\rho_{n,m}$ with $|n-m| \geq 2$ are of higher order ($ \sim \epsilon^{|m-n|}$) and can be thus disregarded. Under such an assumption, the evolution equation for the coherence amplitude $\rho_{n,n+1}$ reads
\begin{equation}
\frac{d \rho_{n,n+1}}{dt} \simeq iJ(P_n-P_{n+1})-\gamma_n \rho_{n,n+1}
\end{equation}
with a site-dependent damping rate $\gamma_n$ given by
\begin{equation}
\gamma_n=\frac{1}{2} \left( J_n^R+J_n^L+J_{n+1}^R+J_{n-1}^L  \right)
\end{equation}
which is of order $ \sim Q$. Equation (36) can be formally solved yielding
\begin{equation}
\rho_{n,n+1}(t)=\rho_{n,n+1}(0) \exp(- \gamma_nt)+iJ \int_0^t dt' [P_n(t')-P_{n+1}(t')] \exp [-\gamma_n(t-t')].
\end{equation}
After an initial fast transient, occurring on a time scale of order $ \sim 1/Q$, the initial value of coherence is damped and $\rho_{n,n+1}(t)$ is driven by the population difference $(P_n-P_{n+1})$ solely, i.e.
\begin{equation}
\rho_{n,n+1}(t) \simeq  i J \int_0^t dt' [P_n(t')-P_{n+1}(t')] \exp [-\gamma_n(t-t')].
\end{equation}
Focusing on the long-time dynamics of the populations, i.e. after the initial fast thermalization process within each potential valley, $P_n(t)$ varies on a time scale much slower that $ 1 / \gamma_n$, and thus Eq.(39) can be approximated as 
\begin{equation}
\rho_{n,n+1}(t) \simeq \frac{iJ}{\gamma_n} \left[ P_n(t)-P_{n+1}(t) \right].
\end{equation}
Substitution of Eq.(40) into Eq.(35) finally yields the following modified rate equation for the populations $P_n$ 
\begin{equation}
\frac{dP_n}{dt} \simeq  \tilde{J}_{n-1}^R P_{n-1}+ \tilde{J}_{n}^L P_{n+1}+(\tilde{J}_{n}^R+\tilde{J}_{n-1}^L) P_n
\end{equation}
where the effective left/right incoherent hopping rates $\tilde{J}_{n-1}^{L,R}$ are given by
\begin{eqnarray}
\tilde{J}_{n}^L=J_n^L+ \Delta_n=Q \exp(h_n)+ \Delta_n  \\
\tilde{J}_{n}^R=J_n^R+ \Delta_n=Q \exp(-h_n)+ \Delta_n  \
\end{eqnarray}
and where
\begin{equation}
\Delta_n= \frac{2J^2}{\gamma_n}= \frac{4J^2}{J_n^R+J_n^L+J_{n+1}^R+J_{n-1}^L}
\end{equation}
describes the coherent-assisted hopping term correction ($ \Delta _n / Q \sim \epsilon^2$).\\ 
At leading order in $\epsilon$, the local force of the random potential reads
\begin{equation}
F_n \equiv \ln \left( \frac{\tilde{J}_n^L}{\tilde{J}_n^R} \right)=2h_n- \frac{2 \Delta_n}{Q} \sinh h_n
\end{equation}
which displays rather generally a nonvanishing mean value of order $\sim \epsilon^2$. However, whenever the probability density function $f(h)$ of the independent and equally-distributed random variables $\{ h_n \}$ has a zero mean and is symmetric around $h=0$, i.e. for $f(-h)= f(h)$, the mean value of the force, $\overline{F_n}$, vanishes.}

\section{Mean-field analysis}
In the purely dissipative limit $J=0$ and using a mean-field approximation \cite{LS7}, one can derive a closed set of equations describing the temporal evolution of the boson occupation number $n_l= \langle a_l^\dagger a_l \rangle={\rm Tr} (\rho(t) a_l^\dagger a_l)$ at various lattice sites. The evolution equation of the mean value of any time-independent operator $\mathcal{O}$, $\langle \mathcal{O} \rangle= {\rm Tr} (\rho(t) \mathcal{O})$, reads
\begin{equation}
\frac{d \langle \mathcal{O} \rangle}{dt}=i \langle [ H, \mathcal{O}] \rangle+ \frac{1}{2} \sum_l \left( \langle [L_l^\dagger,\mathcal{O} ] L_l \rangle +\langle L_l^\dagger [ \mathcal{O},L_l] \rangle \right).
\end{equation}
For $J=0$ (i.e. $H=0$) and dissipators $L_l$ given by Eq.(6) of the main text, taking $\mathcal{O}=a_{l}^{\dagger} a_l$ from Eq.(46) one obtains
\begin{equation}
\frac{dn_l}{dt}  =   (J_l^L-J_l^R) \langle a^\dagger_{l+1}a_{l+1}a^\dagger_l a_l \rangle +(J_{l-1}^R-J_{l-1}^L)
 \langle a^\dagger_l a_l a^\dagger_{l-1}a_{l-1} \rangle -(J^L_{l-1}+J_l^R) n_l+J_{l-1}^Rn_{l-1}+J_l^L n_{l+1} 
 \end{equation}
which is an exact equation. Since the incoherent processes are nonlinear, a closed set of equations for the mean occupation numbers cannot be derived when hopping rates are asymmetric ($J_n^L \neq J_n^R$). However, one can resort to a mean-field approximation by factorizing expectation values as \cite{LS7} $\langle a^\dagger_{l+1}a_{l+1}a^\dagger_l a_l \rangle \simeq \langle a^\dagger_{l+1}a_{l+1}\rangle \langle a^\dagger_l a_l \rangle = n_{l+1}n_l$. This yields the set of closed nonlinear coupled equations
\begin{eqnarray}
\frac{dn_l}{dt}  & = &  (J_l^L-J_l^R) n_{l+1}n_l  +(J_{l-1}^R-J_{l-1}^L)
n_l n_{l-1} -(J^L_{l-1}+J_l^R) n_l+J_{l-1}^Rn_{l-1}+J_l^L n_{l+1} \nonumber \\
& = & \tilde{J}^L_l n_{l+1}+ \tilde{J}_{l-1}^Rn_{l-1}-(J_l^R+J_{l-1}^L)n_l
 \end{eqnarray}
where we have introduced the density-dependent non-linear left/right hopping rates
\begin{equation}
\tilde{J}_l^L=J_l^L+(J_l^L-J_l^R)n_l \; ,\;\;\; \tilde{J}_l^R=J_l^R+(J_l^R-J_l^L)n_{l+1}.
\end{equation}
For a diluite boson distribution with low mean occupation number, i.e. for $n_l \ll 1$, the nonlinear contribution to the hopping rates in Eq.(49) can be neglected and equation (48) reduces to the classical master equation (23) with $n_l=P_l$, and thus the many-particle state displays the Sinai subdiffusive spreading as in the single-particle regime.


\begin{thebibliography}{99}

\bibitem{NH1} Ashida Y, Gong Z and Ueda M 2020 Non-Hermitian physics \emph{Adv. Phys.} \textbf{69} 249
\bibitem{NH2}
Feng L, El-Ganainy R and Ge L 2017 Non-Hermitian photonics based on parity--time symmetry \emph{Nat. Photonics} \textbf{11} 752
\bibitem{NH3}
Longhi S 2018 Parity--time symmetry meets photonics: A new twist in non-Hermitian optics \emph{EPL} \textbf{120} 64001
\bibitem{NH4}
Midya B, Zhao H and Feng L 2018 Non-Hermitian photonics promises exceptional topology of light \emph{Nat. Commun.} \textbf{9} 2674
\bibitem{NH5} Foa Torres L E F 2019 Perspective on topological states of non-Hermitian lattices \emph{J. Phys. Mater.} \textbf{3} 014002
\bibitem{NH6}
Bergholtz E J, Budich J C and Kunst F K 2021 Exceptional topology of non-Hermitian systems \emph{Rev. Mod. Phys.} \textbf{93} 015005
\bibitem{NH7} Zhang X, Zhang T, Lu M H and Chen Y F 2022 A review on non-Hermitian skin effect \emph{Adv. Phys.: X} \textbf{7} 2109431
\bibitem{NH8} Ding K, Fang C and Ma G 2022 Non-Hermitian topology and exceptional-point geometries \emph{Nat. Rev. Phys.} \textbf{4} 74
\bibitem{NH9}
Roccati F, Palma G M, Ciccarello F and Bagarello F 2022 Non-Hermitian physics and master equations \emph{Open Syst. Inf. Dyn.} \textbf{29} 2250004
\bibitem{NH10} Okuma N and Sato M 2023 Non-Hermitian topological phenomena: a review \emph{Annu. Rev. Condens. Matter Phys.} \textbf{14} 83
\bibitem{NH11} Lin R, Tai T, Li L and Lee C H 2023 Topological non-Hermitian skin effect \emph{Front. Phys.} \textbf{18} 53605
\bibitem{NH12} Banerjee A, Sarkar R, Dey S and Narayan A 2023 Non-Hermitian topological phases: principles and prospects \emph{J. Phys.: Condens. Matter} \textbf{35} 333001
\bibitem{NH13} Gohsrich J T, Banerjee A and Kunst F K 2025 The non-Hermitian skin effect: a perspective \emph{EPL} \textbf{150} 60001
\bibitem{NH14}
Xiao L, Wang K, Qu D, Gao H, Lin Q, Bian Z, Zhan X and Xue P 2025 Non-Hermitian physics in photonic systems \emph{Photonics Insights} \textbf{4} R09


\bibitem{NH15}
Gong Z, Ashida Y, Kawabata K, Takasan K, Higashikawa S and Ueda M 2018 Topological phases of non-Hermitian systems \emph{Phys. Rev. X} \textbf{8} 031079
\bibitem{NH16}
Kawabata K, Shiozaki K, Ueda M and Sato M 2019 Symmetry and topology in non-Hermitian physics \emph{Phys. Rev. X} \textbf{9} 041015




\bibitem{SK1} Yao S and Wang Z 2019 Edge states and topological invariants of non-Hermitian systems \emph{Phys. Rev. Lett.} \textbf{121} 086803
\bibitem{SK2} Lee C H and Thomale R 2019 Anatomy of skin modes and topology in non-Hermitian systems \emph{Phys. Rev. B} \textbf{99} 201103
\bibitem{SK3} Kunst F K, Edvardsson E, Budich J C and Bergholtz E J 2018 Biorthogonal bulk-boundary correspondence in non-Hermitian systems \emph{Phys. Rev. Lett.} \textbf{121} 026808
\bibitem{SK4} Yokomizo K and Murakami S 2019 Non-Bloch band theory of non-Hermitian systems \emph{Phys. Rev. Lett.} \textbf{123} 066404
\bibitem{SK5}
Yao S, Song F and Wang Z 2018 Non-Hermitian Chern bands \emph{Phys. Rev. Lett.} \textbf{121} 136802
\bibitem{SK6} Longhi S 2019 Probing non-Hermitian skin effect and non-Bloch phase transitions \emph{Phys. Rev. Research} \textbf{1} 023013
\bibitem{SK7}
Song F, Yao S and Wang Z 2019 Non-Hermitian topological invariants in real space \emph{Phys. Rev. Lett.} \textbf{123} 246801
\bibitem{SK8} Song F, Yao S and Wang Z 2019 Non-Hermitian skin effect and chiral damping in open quantum systems \emph{Phys. Rev. Lett.} \textbf{123} 170401
\bibitem{SK9}
Longhi S 2019 Non-Bloch PT symmetry breaking in non-Hermitian photonic quantum walks \emph{Opt. Lett.} \textbf{44} 5804
\bibitem{SK10}
Lee C H, Li L and Gong J 2019 Hybrid higher-order skin-topological modes in nonreciprocal systems \emph{Phys. Rev. Lett.} \textbf{123} 016805
\bibitem{SK11} Longhi S 2020 Unraveling the non-Hermitian skin effect in dissipative systems \emph{Phys. Rev. B} \textbf{102} 201103(R)
\bibitem{SK12}
Helbig T, Hofmann T, Imhof S, Abdelghany M, Kiessling T, Molenkamp L W, Lee C H, Szameit A, Greiter M and Thomale R 2020 Generalized bulk--boundary correspondence in non-Hermitian topolectrical circuits \emph{Nat. Phys.} \textbf{16} 747
\bibitem{SK13}
Lee C H and Longhi S 2020 Ultrafast and anharmonic Rabi oscillations between non-Bloch bands \emph{Commun. Phys.} \textbf{3} 147
\bibitem{SK14} Borgnia D S, Kruchkov A J and Slager R J 2020 Non-Hermitian boundary modes and topology \emph{Phys. Rev. Lett.} \textbf{124} 056802
\bibitem{SK15} Okuma N, Kawabata K, Shiozaki K and Sato M 2020 Topological origin of non-Hermitian skin effects \emph{Phys. Rev. Lett.} \textbf{124} 086801
\bibitem{SK16} Zhang K, Yang Z and Fang C 2020 Correspondence between winding numbers and skin modes in non-Hermitian systems \emph{Phys. Rev. Lett.} \textbf{125} 126402
\bibitem{SK17} Li L, Lee C H, Mu S and Gong J 2020 Critical non-Hermitian skin effect \emph{Nat. Commun.} \textbf{11} 5491
\bibitem{SK18}
Roccati F 2021 Non-Hermitian skin effect as an impurity problem \emph{Phys. Rev. A} \textbf{104} 022215
\bibitem{SK18b}
Longhi S 2020
Stochastic non-Hermitian skin effect
\emph{Opt. Lett.} \textbf{45} 5250--5253
\bibitem{SK19} Xiao L \emph{et al.} 2020 Non-Hermitian bulk-boundary correspondence in quantum dynamics \emph{Nat. Phys.} \textbf{16} 761
\bibitem{SK20} Yang Z, Zhang K, Fang C and Hu J 2020 Non-Hermitian bulk-boundary correspondence and auxiliary generalized Brillouin zone theory \emph{Phys. Rev. Lett.} \textbf{125} 226402
\bibitem{SK21} Ghatak A, Brandenbourger M, van Wezel J and Coulais C 2020 Observation of non-Hermitian topology and its bulk-edge correspondence in an active mechanical metamaterial \emph{Proc. Natl. Acad. Sci. USA} \textbf{117} 29561
\bibitem{SK22}
Zou D, Chen T, He W, Bao J, Lee C H, Sun H and Zhang X 2021 Observation of hybrid higher-order skin-topological effect in non-Hermitian topolectrical circuits \emph{Nat. Commun.} \textbf{12} 7201
\bibitem{SK23}
Claes J and Hughes T L 2021 Skin effect and winding number in disordered non-Hermitian systems \emph{Phys. Rev. B} \textbf{103} L140201
\bibitem{Brunelli}
Wanjura C C, Brunelli M and Nunnenkamp A 2021
Correspondence between Non-Hermitian Topology and Directional Amplification in the Presence of Disorder
\emph{Phys. Rev. Lett.} \textbf{127} 213601
\bibitem{SK24} Zhang K, Yang Z and Fang C 2022 Universal non-Hermitian skin effect in two and higher dimensions \emph{Nat. Commun.} \textbf{13} 2496
\bibitem{SK25}
Roccati F, Lorenzo S, Calaj\`o G, Palma G M, Carollo A and Ciccarello F 2022 Exotic interactions mediated by a non-Hermitian photonic bath \emph{Optica} \textbf{9} 565
\bibitem{SK26}
Xue W T, Hu Y M, Song F and Wang Z 2022 Non-Hermitian edge burst \emph{Phys. Rev. Lett.} \textbf{128} 120401
\bibitem{SK26b}
Molignini P, Arandes O and Bergholtz E J 2023
Anomalous skin effects in disordered systems with a single non-Hermitian impurity
\emph{Phys. Rev. Res.} \textbf{5} 033058
\bibitem{Brunelli2}
Brunelli M, Wanjura C C and Nunnenkamp A 2023
Restoration of the non-Hermitian bulk-boundary correspondence via topological amplification
\emph{SciPost Phys.} \textbf{15} 173
\bibitem{SK27}
Roccati F, Bello M, Gong Z, Ueda M, Ciccarello F, Chenu A and Carollo A 2024 Hermitian and non-Hermitian topology from photon-mediated interactions \emph{Nat. Commun.} \textbf{15} 2400
\bibitem{SK28}
Xue P, Lin Q, Wang K, Xiao L, Longhi S and Yi W 2024 Self acceleration from spectral geometry in dissipative quantum-walk dynamics \emph{Nat. Commun.} \textbf{15} 4381
\bibitem{SK29}
Wang H Y, Song F and Wang Z 2024 Amoeba formulation of non-Bloch band theory in arbitrary dimensions \emph{Phys. Rev. X} \textbf{14} 021011
\bibitem{SK30}
Xiao L, Xue W T, Song F, Hu Y M, Yi W, Wang Z and Xue P 2024 Observation of non-Hermitian edge burst in quantum dynamics \emph{Phys. Rev. Lett.} \textbf{133} 070801
\bibitem{SK30b}
Midya B 2024 Topological phase transition in fluctuating imaginary gauge fields \emph{Phys. Rev. A} \textbf{109} L061502
\bibitem{SK31}
Longhi S 2024 Incoherent non-Hermitian skin effect in photonic quantum walks \emph{Light: Sci. Appl.} \textbf{13} 95
\bibitem{SK32} Longhi S 2025 Erratic non-Hermitian Skin Localization \emph{Phys. Rev. Lett.} \textbf{134} 196302
\bibitem{SK32b}
Longhi S 2021 Spectral deformations in non-Hermitian lattices with disorder and skin effect: A solvable model \emph{Phys. Rev. B} \textbf{103} 144202

\bibitem{SK33}
Wang S, Wang B, Liu C, Qin C, Zhao L, Liu W, Longhi S and Lu P 2025 Nonlinear non-Hermitian skin effect and skin solitons in temporal photonic feedforward lattices \emph{Phys. Rev. Lett.} \textbf{134} 243805

\bibitem{SK34}
Cai Z{-}F, Li Y, Zhang Y{-}R, Wei X, Yang Z, Liu T and Nori F 2025
Arbitrary control of non-Hermitian skin modes via disorder and an electric field
\emph{arXiv}:2511.16393



\bibitem{uff1}
Metelmann A and Clerk A A 2015 Nonreciprocal photon transmission and amplification via reservoir engineering \emph{Phys. Rev. X} \textbf{5} 021025
\bibitem{uff2}
Metelmann A and T\"ureci H E 2018 Nonreciprocal signal routing in an active quantum network \emph{Phys. Rev. A} \textbf{97} 043833
\bibitem{uff3}
Porras D and Fernandez-Lorenzo S 2019 Topological amplification in photonic lattices \emph{Phys. Rev. Lett.} \textbf{122} 143901
\bibitem{uff4}
Wanjura C C, Brunelli M and Nunnenkamp A 2020 Topological framework for directional amplification in driven-dissipative cavity arrays \emph{Nat. Commun.} \textbf{11} 3149
\bibitem{uff5}
McDonald A, Hanai R and Clerk A A 2022 Nonequilibrium stationary states of quantum non-Hermitian lattice models \emph{Phys. Rev. B} \textbf{105} 064302
\bibitem{LS2}
 Yang F, Jiang Q-D and Bergholtz E J 2022 Liouvillian skin effect in an exactly solvable model \emph{Phys. Rev. Research} \textbf{4} 023160
\bibitem{uff6}
Chaduteau A, Lee D K K and Schindler F 2026
Lindbladian versus Postselected non-Hermitian Topology
\emph{Phys. Rev. Lett.} \textbf{136} 016603

\bibitem{Hatano}
Hatano N and Nelson D R 1996 Localization transitions in non-Hermitian quantum mechanics \emph{Phys. Rev. Lett.} \textbf{77} 570

\bibitem{self}
Longhi S 2022 Non-Hermitian skin effect and self-acceleration \emph{Phys. Rev. B} \textbf{105} 245143




\bibitem{LS1} 
Haga T, Nakagawa M, Hamazaki R and Ueda M 2021 Liouvillian skin effect: slowing down of relaxation processes without gap closing \emph{Phys. Rev. Lett.} \textbf{127} 070402

\bibitem{LS11}
Temme K, Wolf M M and Verstraete F 2012
Stochastic exclusion processes versus coherent transport
\emph{New J. Phys.} \textbf{14} 075004
\bibitem{LS0}
Essler F H L and Piroli L 2020 Integrability of one-dimensional Lindbladians from operator-space fragmentation \emph{Phys. Rev. E} \textbf{102} 062210

 \bibitem{LS3}
Wang Z, Lu Y, Peng Y, Qi R, Wang Y and Jie J 2023 Accelerating relaxation dynamics in open quantum systems with Liouvillian skin effect \emph{Phys. Rev. B} \textbf{108} 054313
\bibitem{LS4}
 Sannia A, Giorgi G L, Longhi S and Zambrini R 2025 Liouvillian skin effect in quantum neural networks \emph{Optica Quantum} \textbf{3} 189-194
\bibitem{LS5}
Cai D-H, Yi W and Dong C-X 2025 Optical pumping through the Liouvillian skin effect \emph{Phys. Rev. B} \textbf{111} L060301
\bibitem{LS6}
 Mao L, Yang X, Tao M-J, Hu H and Pan L 2024 Liouvillian skin effect in a one-dimensional open many-body quantum system with generalized boundary conditions \emph{Phys. Rev. B} \textbf{110} 045440
\bibitem{LS7}
Garbe L, Minoguchi Y, Huber J and Rabl P 2024 The bosonic skin effect: Boundary condensation in asymmetric transport \emph{SciPost Phys.} \textbf{16} 029 
 \bibitem{LS8}
Solanki P, Cabot A, Brunelli M, Carollo F, Bruder C and Lesanovsky I 2025 Generation of entanglement and nonstationary states via competing coherent and incoherent bosonic hopping \emph{Phys. Rev. A} \textbf{112} L030601
 \bibitem{LS9}
Longhi S 2026 Quantum Pontus-Mpemba effect enabled by the Liouvillian skin effect \emph{J. Phys. A: Math. Theor.} \textbf{59}  065304

\bibitem{LS10}
Zhang X, Sun C and Li F 2026 Engineering quantum Mpemba effect by Liouvillian skin effect \emph{arXiv}:2601.16002

 
 \bibitem{T1}
Zhong J X, Kim J W, Longhi S and Jing Y 2026 Observation of erratic non-Hermitian skin localization and transport \emph{arXiv}:2601.19749
 \bibitem{T2}
Nan G, Li Z, Mei F and Xu Z 2026 Anomalous localization and mobility edges in non-Hermitian quasicrystals with disordered imaginary gauge fields \emph{arXiv}:2601.14754
\bibitem{T3}
Miao Y, Ding W, Wang L, Zhao X, Liu S and Yi X 2026 Imaginary gauge-steerable edge modes in non-Hermitian Aubry-Andre-Harper model \emph{arXiv}:2601.06746

 
 \bibitem{Sinai1} 
 Sinai Y G 1982 The limiting behavior of a one-dimensional random walk in a random medium \emph{Theor. Probab. Appl.} \textbf{27} 247
 \bibitem{Sinai2} 
 Bouchaud J P, Georges A 1990 Anomalous diffusion in disordered media: Statistical mechanisms, models and physical applications \emph{Phys. Rep.} \textbf{195} 127
\bibitem{Sinai3} 
Stephen M J 1999 Some results on Sinai diffusion \emph{J. Stat. Phys.} \textbf{96} 1

{\color{black}
\bibitem{Sinai4} 
Bouchaud J P,  Comtet A, Georges A and  Le Doussal P 1990 
Classical diffusion of a particle in a one-dimensional random force field \emph{Ann. Phys.} \textbf{201} 285
} 




\bibitem{D1}
Longhi S 2026 Lifshitz-like metastability and optimal dephasing in dissipative bosonic lattices \emph{Front. Phys.} \textbf{21} 023201






\bibitem{manybody0}
Hamazaki R, Kawabata K and Ueda M 2019
Non-Hermitian many-body localization
\emph{Phys. Rev. Lett.} \textbf{123} 090603

\bibitem{manybody0a}
Kawabata K, Shiozaki K and Ryu S 2022
Many-body topology of non-Hermitian systems
\emph{Phys. Rev. B} \textbf{105} 165137


\bibitem{manybody0b}
Zhang W, Di F, Yuan H, Wang H, Zheng X, He L, Sun H and Zhang X 2022
Observation of non-Hermitian aggregation effects induced by strong interactions
\emph{Phys. Rev. B} \textbf{105} 195131


\bibitem{manybody1c}
Brighi P and Nunnenkamp A 2024
Nonreciprocal dynamics and the non-Hermitian skin effect of repulsively bound pairs
\emph{Phys. Rev. A} \textbf{110} L020201


\bibitem{manybody1b}
Brighi P, Ljubotina M, Roccati F and Balducci F 2025
Finite steady-state current defies non-Hermitian many-body localization
\emph{Phys. Rev. Research} \textbf{7} L042014


\bibitem{manybody1}
Gliozzi J, Balducci F, Hughes T L and De Tomasi G 2025
Non-Hermitian multipole skin effects challenge localization
\emph{arXiv}:2504.10580 [cond-mat.dis-nn]


\bibitem{manybody2}
Brighi P and Nunnenkamp A 2025
Pairing-induced phase transition in the non-reciprocal Kitaev chain
\emph{arXiv}::2510.24851 [quant-ph]

{\color{black}




\bibitem{timescale}
Padash A, Aghion E, Schulz A, Barkai E, Chechkin A V, Metzler R and  Kantz H 2022 
Local equilibrium properties of ultraslow diffusion in the Sinai model \emph{New J. Phys.} \textbf{24} 073026


\bibitem{pert1}
Bernal-Garcia D N,  Rodriguez B A and Vinck-Posada H 2019 Multiple-scale analysis of open quantum systems
\emph{Phys. Lett. A} \textbf{ 383} 1698 

\bibitem{pert2}
Longhi S 2024 Dephasing-Induced Mobility Edges in Quasicrystals \emph{Phys. Rev. Lett.} \textbf{132} 236301

}







\end{thebibliography}
\end{document}